\documentclass{article}

\usepackage[a4paper]{geometry}
\usepackage{amsmath,amssymb,amsthm,mathtools}
\usepackage[colorlinks,citecolor=blue,urlcolor=blue]{hyperref}
\usepackage[affil-it]{authblk}
\usepackage[round]{natbib}
\title{Ranking soccer teams on basis of their current strength: a comparison of maximum likelihood approaches} 
\author{Christophe Ley
\thanks{Electronic address: \texttt{christophe.ley@ugent.be}}}
\affil{Ghent University, Ghent, Belgium}

\author{Tom Van de Wiele %
    \thanks{Electronic address: \texttt{tomvandewiele@google.com}}}
\affil{DeepMind, London, UK}
\author{Hans Van Eetvelde
\thanks{Electronic address: \texttt{hans.vaneetvelde@ugent.be}}}
\affil{Ghent University, Ghent, Belgium}
\date{}

\begin{document}

\maketitle

\begin{abstract}
\indent We present  ten different strength-based statistical models that we use to model soccer match outcomes with the aim of producing a new ranking. The models are of four main types: Thurstone-Mosteller,  Bradley-Terry, Independent Poisson and Bivariate Poisson, and their common aspect is that the parameters are estimated via weighted maximum likelihood, the weights being a match importance factor and a time depreciation factor giving less weight to matches that are played a long time ago. Since our goal is to build a ranking reflecting the teams' current strengths, we compare the 10 models on basis of their predictive performance via the Rank Probability Score at the level of both domestic leagues and national teams. We find that the best models are the Bivariate and Independent Poisson models. We then illustrate the versatility and usefulness of our new rankings by means of three examples where the existing rankings fail to provide enough information or lead to peculiar results.
\end{abstract}


\section{Introduction}
\label{sec:intro} 

Football, or soccer, is undeniably the most popular sport worldwide. Predicting which team will win the next World Cup or the Champions League final are issues that lead to heated discussions and debates among football fans, and even attract the attention of casual watchers. Or put more simply, the question of which team will win the next match, independent of its circumstances, excites the fans. Bookmakers have made a business out of football predictions, and they use highly advanced models taking into account numerous factors (like a team's current form, injured players, the history between both teams, the importance of the game for each team, etc.) to obtain the odds of winning, losing and drawing for both teams. 

One major appeal of football, and a reason for its success, is its simplicity as game. This stands somehow in contrast to the difficulty of predicting the winner of a football match. A help in this respect would be a ranking of the teams involved in a given competition based on their current strength, as this would enable football fans and casual watchers to have a better feeling for who is the favourite and who is the underdog. However, the existing rankings, both at domestic leagues level and at national team level, fail to provide this, either because they are by nature not designed for that purpose or because they suffer from serious flaws.

Domestic league rankings obey the 3-1-0 principle, meaning that the winner gets 3 points, the loser 0 points and a draw earns each team 1 point. The ranking is very clear and fair, and tells at every moment of the season how strong a team has been since the beginning of the season. However, given that every match has the same impact on the ranking, it is not designed to reflect a team's current strength. {A  recent illustration of this fact can be found in last year's English Premier League, where the newly promoted team of Huddersfield Town had a very good start in the  season 2017-2018 with 7 out of 9 points after the first 3 rounds. They ended the first half of the season on rank 11 out of 20, with 22 points after 19 games. Their second half season was however very poor, with only 15 points scored in 19 games, earning them the second last spot over the second half of the season (overall they ended the year on rank 16, allowing them to stay in the Premier League). There was a clear tendency of decay in their performance, which was hidden in the overall ranking by their very good performance at the start of the season.}

Contrary to domestic league rankings, the FIFA/Coca-Cola  World Ranking of national soccer teams is intended to rank teams according to their recent performances in international games. Bearing in mind that the FIFA ranking forms the basis of the seating and the draw in international competitions and its qualifiers, such a requirement on the ranking is indeed necessary. However, the current FIFA ranking\footnote{While the present paper was in the final stages of the revision procedure, the FIFA decided to change its ranking in order to avoid precisely the flaws we mention here. Given the short time constraint, we were not able to study their new ranking and leave this for future research.} fails to reach these goals in a satisfying way and is subject to many discussions (\citet{Cummings,Tweedale, TAP}). It is  based on the 3-1-0 system, but each match outcome is multiplied by several factors like the opponent team's ranking and confederation, the importance of the game, and a time factor. We spare the reader those details here, which can be found on the webpage of the FIFA/Coca-Cola World Ranking (\url{https://www.fifa.com/mm/document/fifafacts/rawrank/ip-590_10e_wrpointcalculation_8771.pdf}).   In brief, the  ranking  is based on the weighted average of ranking points  a national team has won over each of the preceding four rolling years.  The average ranking points over the last 12 month period make up half of the ranking points, while the average ranking points in the 13-24 months before the update count for 25\% leaving 15\% for the 25-36 month period  and 10\% for the 37-48 month period before the update. This {arbitrary decay function} is a  major criticism of the FIFA ranking: a similar match of eleven months ago can have approximately twice the contribution as a match played twelve months ago. A striking example hereof was Scotland: ranked $50^{\rm th}$ in August 2013, it dropped to rank 63 in September 2013 before making a major jump to rank 35 in October 2013. This high volatility demonstrates a clear weakness in the FIFA ranking's ability of mirroring a team's current strength.

In this paper, we intend to fill the gap and develop  a ranking that does reflect a soccer team's current strength. To this end, we consider and compare various existing and new statistical models that assign one or more strength parameters to each soccer team and where these parameters are estimated over an entire range of matches by means of maximum likelihood estimation. We shall propose a smooth time depreciation function to give more weight to more recent matches. {The comparison between the distinct models will be based on their predictive performance, as the model with the best predictive performance will also yield the best current-strength-ranking.} The resulting ranking represents an interesting addition to the well-established rankings of domestic leagues and can be considered as promising alternative to the FIFA ranking of national teams.

The present paper is organized as follows. We shall present in Section~\ref{sec:models}  10 different strength-based models whose parameters are estimated via maximum likelihood. More precisely, via weighted maximum likelihood as we introduce two types of weight parameters: the above-mentioned time depreciation effect and a match importance effect for national team matches. In Section~\ref{statistics} we describe the exact computations behind our estimation procedures as well as {a criterion} according to which we define a statistical model's predictive performance. Two case studies allow us to compare our 10 models at domestic league and national team levels in Section~\ref{sec:comp}: we investigate the English Premier League seasons from 2008-2017 (Section~\ref{sec:PL}) as well as  national team matches between 2008 and 2017 (Section~\ref{sec:NT}). On basis of the best-performing models, we  then  illustrate in Section~\ref{sec:rankings} the advantages of our current-strength based ranking via various examples.  We conclude the paper with final comments and an outlook on future research in Section~\ref{sec:conclu}.

\section{The statistical strength-based models}\label{sec:models}

\subsection{Time depreciation and match importance factors}\label{sec:weights}

Our strength-based statistical models are of two main types: Thurstone-Mosteller and  Bradley-Terry type models on the one hand, which directly model the outcome (home win, draw, away win) of a match, and the Independent and Bivariate Poisson models on the other hand, which model the scores of a match. Each model assigns strength parameters to all teams involved and models match outcomes via these parameters. Maximum likelihood  estimation is employed to estimate the strength parameters, and the teams are  ranked  according to their resulting overall strengths. More precisely, we shall consider weighted maximum likelihood estimation, where the weights introduced are of two types: time depreciation (domestic leagues and national teams) and match importance (only national teams).

\subsubsection{A smooth decay function based on the concept of Half period}

\begin{figure}
\begin{center}
\includegraphics[width=0.8\linewidth]{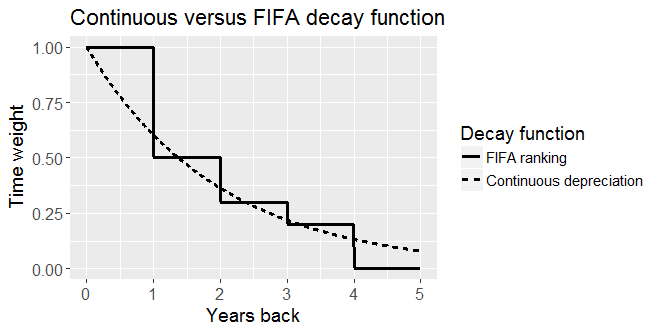}
\end{center}
\caption{Comparison of the FIFA ranking decay function versus our exponential smoother~\eqref{smoother}. The  continuous depreciation line uses a Half Period of 500 days.}
\label{fig:decay}
\end{figure}

A feature that is common to all considered models is our proposal of decay function in order to reflect the time depreciation. Instead of the step-wise decay  function employed in the FIFA ranking, we rather suggest a continuous depreciation function that gives less weight to older matches with a maximum weight of 1 for a match played today. Specifically, the time weight for a match which is played $x_m$ days back is calculated as 
\begin{equation}\label{smoother}
w_{time,m}(x_m) = \left(\frac{1}{2}\right)^{\frac{x_m}{\mbox{Half period}}},
\end{equation}
meaning that a match played \emph{Half period} days ago only contributes half as much as a match played today and a match played $3\times$\emph{Half period} days ago contributes 12.5\% of a match played today. Figure~\ref{fig:decay} shows a graphical comparison of our continuous time decay function versus the arbitrary FIFA decay function. In the sequel,  $w_{time,m}$ will serve as weighting function in the likelihoods associated with our various models. This idea of weighted likelihood or pseudo-likelihood to better estimate a team's current strength is in line with the literature on modelling (mainly league) football scores, see~\citet{dixon1997modelling}.

\subsubsection{Match importance}

While in domestic leagues all matches are equally important, the same cannot be said about national team matches where for instance friendly games are way less important than matches played during the World Cup. Therefore we need to introduce importance factors. The FIFA weights seem reasonable for this purpose and will be employed whenever national team matches are analyzed. The relative importance of a national match is indicated by $w_{type,m}$ and can take the values 1 for a friendly game, 2.5 for a confederation or world cup qualifier, 3 for a confederation tournament (e.g., UEFA EURO2016 or the Africa Cup of Nations 2017) or the confederations cup, and 4 for World Cup matches. 

\subsection{The Thurstone-Mosteller and Bradley-Terry type models}

Thurstone-Mosteller (TM) \citep{thurstone1927psychophysical,mosteller2006remarks} and Bradley-Terry (BT) models \citep{BT52} have been designed to predict the outcome of pairwise comparisons. Assume from now on that we look at $M$ matches involving in total $T$ teams. Both models consider latent continuous variables $Y_{i,m}$ which stand for the performance of team $i$ in match $m$, $i \in \{1,\ldots,T\}$ and $m \in \{1,\ldots,M\}$. When the performance of team $i$ is much better than the performance of team $j$ in match $m$, say $Y_{i,m}-Y_{j,m}>d$ for some positive real-valued $d$, then team $i$ beats team $j$ in that match. If the difference in their performances is lower than $d$,  i.e. $|Y_{i,m}-Y_{j,m}|<d$, then the game will end in a draw. The parameter $d$ thus determines the overall chance for a draw. The performances $Y_{i,m}$ depend on the strengths of the teams, denoted by $r_i$ for $i \in \{1,\ldots,T\}$, implying that a total of $T$ team strengths need to be estimated.

\subsubsection{Thurstone-Mosteller model}

The Thurstone-Mosteller model assumes that the performances $Y_{i,m}$ are normally distributed with means $r_{i}$, the strengths of the teams. The variance is considered to be the same for all teams, which leads to $Y_{i,m}\sim N(r_i, \sigma^2)$. Since the variance $\sigma^2$ only determines the scale of the ratings $r_i$, it can be chosen arbitrarily. Another assumption is that the performances of teams are independent, implying that $Y_{i,m}-Y_{j,m}\sim N(r_i-r_j, 2\sigma^2)$. For games not played on neutral ground, a parameter $h$ is added to the strength of the home team. In the remainder of this article, we will assume that team $i$ is the home team and has the home advantage, unless stated otherwise.

If we call $P_{H_{ijm}}$ the probability of a home win in match $m$, $P_{D_{ijm}}$ the probability of a draw in match $m$ and $P_{A_{ijm}}$ the probability of an away win in match $m$, then the outcome probabilities are
\begin{align*} 
P_{H_{ijm}} &=P(Y_{i,m}-Y_{j,m}>d)= \Phi\left(\frac{(r_{i}+h)-r_{j}-d}{\sigma \sqrt{2}}\right); \\ 
P_{A_{ijm}} &=P(Y_{j,m}-Y_{i,m}>d)= \Phi\left(\frac{r_{j}-(r_{i}+h)-d}{\sigma \sqrt{2}}\right);\\
P_{D_{ijm}} &=  1-P_{H_{ijm}}-P_{A_{ijm}}, 
\end{align*} 
where $\Phi$ denotes the cumulative distribution function of the standard normal distribution. For the sake of clarity we wish to stress that $r_{i}$ and $r_{j}$ belong to the set $\{r_1,\ldots,r_T\}$ of all $T$ team strengths. In principle we should adopt the notation $r_{i(m)}$ and $r_{j(m)}$ with $i(m)$ and $j(m)$ indicating the home and away team in match $m$; however, we believe this notation is too heavy and the reader readily understands what we mean without these indices. If the home effect $h$ is greater than zero, it inflates the strength of the home team and increases its modeled probability to win the match. This is typically the case since playing at home gives the benefit of familiar surroundings, the support of the home crowd and the lack of traveling. Matches on neutral ground are modeled by dropping the home effect $h$.

The strength parameters  are estimated using maximum likelihood estimation on  match outcomes. Let $y_{R_{ijm}}$ be 1 if the result of match $m$ is $R$ and $y_{R_{ijm}} = 0$ otherwise, for $R=H,D,A$ as explained above. Under the common assumption  that matches are independent, the likelihood for $M$ matches corresponds to  
\begin{align} 
L &= \prod_{m=1}^{M}\prod_{i,j\in\{1,\ldots,T\}}\prod_{R\in \{H,D,A\}}P_{R_{ijm}}^{y_{ijm}\cdot y_{R_m}\cdot w_{type,m} \cdot w_{time,m}} \label{likelihood}
\end{align}
with $w_{type,m}$ and $w_{time,m}$ the weights described in Section~\ref{sec:weights} and where $y_{ijm}$ equals 1 if $i$ and $j$ are the home resp. the away team in match $m$ and $y_{ijm}=0$ otherwise.

\subsubsection{Bradley-Terry model}

In the Bradley-Terry model, the normal distribution is replaced with the logistic distribution. This leads to the assumption  that $Y_{i,m}-Y_{j,m}\sim logistic(r_i-r_j, s)$ where again the scale parameter $s$ is considered equal for all teams and can be chosen arbitrarily. The corresponding outcome probabilities are
\begin{align*} 
P_{H_{ijm}} &=P(Y_{i,m}-Y_{j,m}>d)=\frac{1}{1+\exp\left(-\frac{(r_{i}+h)-r_{j}-d}{s}\right)} ; \\ 
P_{A_{ijm}} &=P(Y_{j,m}-Y_{i,m}>d)=\frac{1}{1+\exp\left(-\frac{r_{j}-(r_{i}+h)-d}{s}\right)} ;\\
P_{D_{ijm}} &=  1-P_{H_{ijm}}-P_{A_{ijm}}, 
\end{align*} 
where again $h$ and $d$ stand for the home effect parameter and draw parameter  and $r_{i}$ and $r_{j}$ respectively stand for  the strength parameters of home and away team in match $m$. The parameters are estimated via maximum likelihood in the same way as for the Thurstone-Mosteller model.

\subsubsection{Bradley-Terry-Davidson model}

In the original Bradley-Terry model,  there exists no possibility for a draw ($d=0$). The two possible outcomes can then be written in a very simple and easy-to-understand formula, if we transform the parameters by taking $r_i^*=\exp(r_i/s)$ and $h^*=\exp(h/s)$:
\begin{align*} 
P_{H_{ijm}} &=\frac{h^* r_i^*}{h^* r_i^*+r_j^*} ;\\
P_{A_{ijm}} &=\frac{r_j^*}{h^* r_i^*+r_j^*}.  
\end{align*} 
These simple formulae are one of the reasons for the popularity of the Bradley-Terry model.  Starting from there, \citet{Davidson} modeled the draw probability in the following way:
\begin{align*} 
P_{H_{ijm}} &=\frac{h^* r_i^*} {h^*r_i^*+d^*\sqrt{h^*r_i^*r_j^*} +r_j^*} ;\\
P_{A_{ijm}} &=\frac{r_j^*}{h^*r_i^*+d^*\sqrt{h^*r_i^*r_j^*} +r_j^*} ;\\  
P_{D_{ijm}} &=\frac{d^*\sqrt{h^*r_i^*r_j^*}}{h^*r_i^*+d^*\sqrt{h^*r_i^*r_j^*} +r_j^*}.
\end{align*} 
The draw effect $d^*$ is best understood by assuming similar strengths in the absence of a home effect. In that case $P_{H_{ijm}}$ is similar to $P_{A_{ijm}}$ and the relative probability of $P_{D_{ijm}}$ compared to a home win or loss is approximately equal to $d^*$. Parameter estimation works in the same way as in the previous two sections.

\subsubsection{Thurstone-Mosteller,  Bradley-Terry and Bradley-Terry-Davidson models with Goal Difference weights}

The basic Thurstone-Mosteller,  Bradley-Terry and Bradley-Terry-Davidson models of the previous sections do not use all of the available information. They only take the match outcome into account, omitting likely valuable information present in the goal difference. A team that wins by 8-0 and loses the return match by 0-1 is probably stronger than the opponent team. Therefore we propose an extension of these models that modifies the basic models in the sense that matches are given an increasing weight when the goal difference grows. The likelihood function is calculated as
\begin{align*} 
L &= \prod_{m=1}^{M}\prod_{i,j\in\{1,\ldots,T\}}\prod_{R\in\{H,D,A\}}P_{R_{ijm}}^{y_{ijm}*y_{R_{ijm}}\cdot w_{goalDiffscaled,m} \cdot w_{type,m} \cdot w_{time,m}},
\end{align*}
where  $P_{R_{ijm}}$ can stand for the Thurstone-Mosteller,  Bradley-Terry and Bradley-Terry-Davidson expressions respectively, leading to three new models. This formula slightly differs from~\eqref{likelihood} through the goal difference weight 
$$
w_{goalDiffscaled,m}=\left\{\begin{array}{ll}
1&\mbox{if draw}\\
\log_2(goalDiff_m+1)&\mbox{else,}
\end{array}\right.
$$ 
with $goalDiff_m$ the absolute value of the goal difference in match $m$ (both outcomes 2-0 and 0-2 thus give the same goal difference of 2). This way, a goal difference of 1 receives a goal difference weight of 1 and every additional increment in goal difference results in a smaller increase of the goal difference weight. A goal difference of 7 goals receives a goal difference weight of 3. Parameter estimation is achieved in the same way as in the basic  models.

\subsection{The Poisson models}

Poisson models were first suggested by \cite{PoissonMaher} to model football match results. He assumed the number of scored goals by both teams to be {independent} Poisson distributed variables. Let $G_{i,m}$ and $G_{j,m}$ be the random variables representing the goals scored by team $i$ and team $j$ in match $m$, respectively. With those assumptions the probability  function can be written as
\begin{align} 
{\rm P}(G_{i,m}=x,G_{j,m}=y) &= \frac{\lambda_{i,m}^x}{x!}\exp(-\lambda_{i,m}) \cdot \frac{\lambda_{j,m}^y}{y!}\exp(-\lambda_{j,m}), \label{poissonDens}
\end{align}
where $\lambda_{i,m}$ and $\lambda_{j,m}$ stand for  the means of $G_{i,m}$ and $G_{j,m}$, respectively. In what follows we shall consider this model and variants of it, including the Bivariate Poisson model that removes the independence assumption.

Being a count-type distribution, the Poisson is a natural choice to model soccer matches. It bares yet another advantage when it comes to predicting matches. If $GD_m = G_{i,m} - G_{j,m}$, then the probability
of a win of team $i$ over team $j$,  the probability of a draw as well as the win of team $j$ in match $m$ are respectively computed as ${\rm P}(GD_m > 0)$, ${\rm P}(GD_m = 0)$ and ${\rm P}(GD_m<0)$. The Skellam distribution, the discrete probability distribution of the difference of two independent Poisson random variables, is used to derive these probabilities given $\lambda_{i,m}$ and~$\lambda_{j,m}$. This renders the prediction of future matches via the Poisson model particularly simple.

\subsubsection{Independent Poisson model}
Attributing again a single strength parameter to each team, {denoted as before by $r_1,\ldots,r_T$, and keeping the notation $r_{i},r_{j}\in\{r_1,\ldots,r_T\}$ for the home and away team strengths in match $m$}, we define the Poisson means as $\lambda_{i,m} = \exp( c +  (r_{i}+h) - r_{j})$ and $\lambda_{j,m} = \exp( c +  r_{j} - (r_{i}+h))$ with $h$  the home effect, $c$  a common intercept.  Matches on neutral ground are modeled by dropping the home effect $h$. With this in hand, the overall likelihood can be written as
$$
L = \prod_{m=1}^{M}\prod_{i,j\in \{1,...,T\}} \left(\frac{\lambda_{i,m}^{g_{i,m}}}{g_{i,m}!}\exp(-\lambda_{i,m}) \cdot \frac{ \lambda_{j,m}^{g_{j,m}}}{g_{j,m}!} \exp(-\lambda_{j,m})\right)^{y_{ijm} \cdot w_{type,m} \cdot w_{time,m}},
$$
where $y_{ijm}=1$ if $i$ and $j$ are the home team, resp. away team in match $m$ and $y_{ijm}=0$ otherwise, and $g_{i,m}$ and $g_{j,m}$ stand for the actual goals made by both teams in match $m$. Maximum likelihood estimation yields the values of the strength parameters. It is important to notice that the Poisson model uses two observations for each match (the goals scored by each team) while using the same number of parameters (number of teams + 2). The TM and BT models, except for the models with Goal Difference Weight, only use a single observation for each match.

\subsubsection{The Bivariate Poisson model}

A potential drawback of the Independent Poisson models lies precisely in the independence assumption. Of course, some sort of dependence between the two playing teams is introduced by the fact that the strength parameters of each team are present in the Poisson means of both teams, however this may not be a sufficiently rich model to cover the interdependence between two teams. 

\cite{PoissonBivariate} suggested a bivariate Poisson model by adding a correlation between the scores. The scores in a match between teams $i$ and $j$ are modelled as $G_{i,m}=X_{i,m}+X_{C}$ and  $G_{j,m}=X_{j,m}+X_{C}$, where $X_{i,m}$, $X_{j,m}$ and $X_{C}$ are independent Poisson distributed variables with parameters $\lambda_{i,m}$, $\lambda_{j,m}$ and $\lambda_{C}$, respectively. The joint probability function of the home and away score is then given by
\begin{align} 
{\rm P}(G_{i,m}=x, G_{j,m}=y)=\frac{\lambda_{i,m}^x \lambda_{j,m}^y}{x!y!} \exp(-(\lambda_{i,m}+\lambda_{j,m}+\lambda_{C})) \sum_{k=0}^{\min(x,y)} \binom{x}{k} \binom{y}{k}k!\left(\frac{\lambda_{C}}{\lambda_{i,m}\lambda_{j,m}}\right)^k, \label{bivpoissonDens}
\end{align}
which is the formula for the bivariate Poisson distribution with parameters $\lambda_{i,m}$, $\lambda_{j,m}$ and $\lambda_{C}$. It reduces to~\eqref{poissonDens} when $\lambda_{C}=0$. This parameter thus can be interpreted as the covariance between the home and away scores in match $i$ and might reflect the game conditions. The means $\lambda_{i,m}$ and $\lambda_{j,m}$ are similar as in the Independent model, but we attract the reader's attention to the fact that the means for the scores are now given by $\lambda_{i,m}+\lambda_{C}$ and $\lambda_{j,m}+\lambda_{C}$, respectively. We assume that the covariance $\lambda_{C}$ is constant  over all matches. All $T+3$ parameters are again estimated by means of maximum likelihood estimation.
 
Letting $GD_m$ again stand for the goal difference,  we can easily see that the probability function of the goal difference for the bivariate case is the same as the probability function for the Independent model with  parameters $\lambda_{i,m}$ and $\lambda_{j,m}$, since
\begin{align*}
    P(GD_m=x)&=P(G_{i,m}-G_{j,m}=x)\\ &= P(X_{i,m}+X_{C}-(X_{j,m}+X_{C})=x) = P(X_{i,m}-X_{j,m}=x),
\end{align*}
implying that we can again use the Skellam distribution for predicting the winner of future games.

One can think of many other ways to model dependent football scores.  \cite{PoissonBivariate} also consider bivariate Poisson models where the dependence parameter $\lambda_C$ depends on either the home team, either the away team, or both teams. We do not include these models here as they are more complicated and, in preliminary comparison studies that we have done, always performed worse than the above-mentioned model with constant $\lambda_C$. Other ways to model the dependence between the home and away scores have been proposed in the literature. For instance, the dependence can be modelled by all kinds of copulas or  adaptations of the Independent model. Incorporating them all in our analysis seems an impossible task, which is why we  opted for the very prominent Karlis-Ntzoufras proposal. Notwithstanding, we mention some important contributions in this field: \cite{dixon1997modelling} added an additional parameter to adjust for the probabilities on low scoring games (0-0, 1-0, 0-1 and 1-1), \cite{HaleScarf}  investigated copula dependence  structures,  and  \cite{boshnakov2017bivariate} recently  proposed  a  copula-based bivariate  Weibull count model.  

\subsubsection{Poisson models with defensive and attacking strengths}
In the previous sections we have defined a slightly simplified version of Maher's original idea. In fact, Maher assumed  the scoring rates to be of the form $\lambda_{i,m} = \exp(c + (o_{i}+h) - d_{j})$ and $\lambda_{j,m} = \exp(c + o_{j} - (d_{i}+h))$, with  $o_{i}$, $o_{j}$, $d_{i}$ and $d_{j}$ standing for offensive and defensive capabilities of teams $i$ and $j$ in match $m$. This allows us to extend both the Independent and Bivariate Poisson model to incorporate offensive and defensive abilities, opening the door to the possibility of an offensive and defensive ranking of the teams. These models thus consider $2T$ team strength parameters to be estimated via maximum likelihood.

Since every team is given two strength parameters in this case, one may wonder how to build rankings. We suggest two options. On the one hand, this model can lead to two rankings, one for attacking strengths and the other for defensive strengths. On the other hand, we can simulate a round-robin tournament with the estimated strength parameters and consider the resulting ranking. We refer the reader to \cite{scarf2011numerical} for details about this approach.

\section{Parameter estimation and model selection}\label{statistics}

In this section we shall briefly describe two crucial statistical aspects of our investigation, namely how we compute the maximum likelihood estimates and which criterion we apply to select the model with the highest predictive performance.

\subsection{Computing the maximum likelihood estimates}

Parameters in the Thurstone-Mosteller and Bradley-Terry type as well as in the Poisson  models are estimated using maximum likelihood estimation. To this end, we have used the $optim$ function in $\mathtt{R}$ \citep{Rteam}  by specifying as preferred method the \textit{BFGS} (Broyden-Fletcher-Goldfarb-Shanno optimization algorithm). We have opted for this quasi-Newton method because of its robust properties. Note that the ratings $r_i$ are unique up to addition by a constant. To identify these parameters, we add the constraint that the sum of the ratings has to equal zero. For the Bradley-Terry-Davidson model the same constraint can be applied after logtransformation of the ratings $r_i^*$. Thanks to this constraint, only $T-1$ strengths have to be estimated when we consider $T$ teams. For the models with 2 parameters per team, we have to estimate $2(T-1)$ strength parameters. The strictly positive parameters are initialized at 1, the other parameters get an initial value of 0. After the first optimization, the estimates are used as initial values in the next optimization to speed up the calculations.

\subsection{Measure of predictive performance}\label{predperf}
The studied models are built to perform three-way outcome prediction (home win, draw or home loss). Each of the three possible match outcomes is predicted with a certain probability but only the actual outcome is observed. The predicted probability of the outcome that was actually observed is thus a natural measure of predictive performance. The ideal predictive performance metric is able to select the model which approximates the true outcome probabilities the best. 

The metric we use is the Rank Probability Score (RPS) of \cite{epstein1969scoring}. It represents the difference between cumulative predicted and observed distributions via the formula
$$
\frac{1}{2M}\sum_{m=1}^M\left((P_{H_m}-y_{H_m})^2+(P_{A_M}-y_{A_M})^2\right)
$$
where we simplify the previous notations so that $P_{H_m}$ and $P_{A_m}$ stand for the predicted probabilities in match $m$ and $y_{H_m}$ and $y_{A_m}$ for the actual outcomes (hence, 1 or 0). It has been shown in \cite{constantinou2012solving} that the RPS is more appropriate as soccer performance metric than other popular metrics such as the RMS and Brier score. The reason is that, by construction, the RPS works at an ordinal instead of nominal scale, meaning that, for instance, it penalizes more severely a wrongly predicted home win in case of a home loss than in case of a draw.

\section{Comparison of the 10 models in terms of their predictive performance}\label{sec:comp}
In this section we compare the predictive performance of all 10 models described in Section~\ref{sec:models}. To this end, we first consider the English Premier League  as example for domestic league matches, and then move to national team matches played over a period of 10 years all over the world, i.e., without restriction to a particular zone.

\subsection{Case study 1: Premier League}\label{sec:PL}

 The engsoccerdata package \citep{engsoccerdata}   contains results of all top 4 tier football leagues in England since 1888. The dataset contains the date of the match, the teams that played, the tier   as well as the result. The number of teams equals 20 for each of the seasons considered (2008-2017). Matches are predicted for every season separately and on every match day  of the season, using two years for training the models. We left out the first 5 rounds of every season, so a total of 3300 matches are predicted. The reason for the burn-in period is the fact that for the new teams in the Premier League, we can not have a good estimation yet of their strength at the beginning of the season since we are lacking information about the previous season(s). Matches are predicted in blocks corresponding to each round, and after every round the parameters are updated. In all our models, the Half Period is varied between 30 days and 2 years in steps of 30 days. 

Table \ref{modelSummaryPL} summarizes the analysis by comparing the best performing models of each of the  considered classes, \mbox{i.e.} the model with the optimal Half Period. As we can see, the Bivariate Poisson model with 1 strength parameter per team is the best according to the RPS, followed by the Independent Poisson model with just one parameter per team. So parsimony in terms of parameters to estimate is important. We also clearly see that all Poisson-based models  outperform the TM and BT type models. This was to be expected since Poisson models use the goals as additional information. Considering the goal difference in the TM and BT type models does not improve their performance. It is also noteworthy that the best two models have among the lowest Half Periods. 

\begin{table}
\caption{\label{modelSummaryPL}Comparison table for the best performing models of each of the  considered classes with respect to the RPS criterion. The English Premier League matches from rounds 6 to 38 between the seasons 2008-2009 and 2017-2018 are considered.} \vspace{.5cm}

\centering 
\begin{tabular}{lll}
\textbf{Model Class} & 
\textbf{Optimal Half Period}&
\textbf{RPS} \\
 Bivariate Poisson                  & 390 days & 0.1953  \\
 Independent Poisson                & 360 days & 0.1954  \\
 Independent Poisson Def. \& Att.   & 390 days & 0.1961  \\
 Bivariate Poisson Def \& Att.      & 480 days & 0.1961 \\
 Thurstone-Mosteller                & 450 days & 0.1985  \\ 
 Bradley-Terry-Davidson             & 420 days & 0.1985  \\
 Bradley-Terry                      & 420 days & 0.1986  \\
 Thurstone-Mosteller + Goal Difference    & 300 days & 0.2000 \\
 Bradley-Terry-Davidson + Goal Difference & 420 days & 0.2000 \\
 Bradley-Terry + Goal Difference    & 450 days & 0.2003 \\
\end{tabular}
\end{table}

\subsection{Case study 2: National teams}\label{sec:NT}

For the national team match results we used the dataset ``International football results from 1872 to 2018" uploaded by Mart J\"urisoo on the website \url{https://www.kaggle.com/}.  We predicted the outcome of 4268 games played all over the world in the period from 2008 to 2017. The last game in our analysis is played on 2017-11-15. To avoid a too extreme computational time, we left out the friendly games in the comparison. The parameters are estimated by maximum likelihood on a period of eight years. The Half Period is varied from a half year to six years in steps of a half year.

The results of our model comparison are provided in Table~\ref{modelSummaryNT}. Exactly as for the Premier League, the Bivariate Poisson model with 1 strength parameter per team comes out first, followed by the Independent Poisson model with 1 strength parameter. We retrieve also all the other conclusions from the domestic level comparison. It is interesting to note that a Half Period of 3 years leads to the lowest RPS for both best models. Given the sparsity of national team matches played over a year, we think that no additional level of detail such as 3 years and 2 months is required, as this may also lead to  over-fitting.

\begin{table}
\caption{\label{modelSummaryNT}Comparison table for the best performing models of each of the  considered classes with respect to the RPS criterion. All of the important matches between the national teams in the period 2008-2017 are considered.}\vspace{.5cm}
\centering 
\begin{tabular}{lll}
\textbf{Model Class} & 
\textbf{Optimal Half Period}&
\textbf{RPS} \\
 Bivariate Poisson                  & 3 years & 0.1651  \\
 Independent Poisson                & 3 years & 0.1653  \\
 Independent Poisson Def. \& Att.   & 3.5 years & 0.1656  \\
 Bivariate Poisson Def \& Att.      & 3 years & 0.1656 \\
 Thurstone-Mosteller                & 3.5 years & 0.1658 \\
 Bradley-Terry                      & 4 years & 0.1659  \\
 Bradley-Terry-Davidson             & 4 years & 0.1660  \\ 
 Thurstone-Mosteller + Goal Difference    & 3.5 years & 0.1672  \\
 Bradley-Terry + Goal Difference    & 3 years & 0.1674 \\
 Bradley-Terry-Davidson + Goal Difference & 3.5 years & 0.1681 \\
\end{tabular}
\end{table}

\section{Applications of our new rankings}\label{sec:rankings}

We now illustrate the usefulness of our new current-strength based rankings by means of various examples. Given the dominance of the Bivariate Poisson model with 1 strength parameter in both settings, we will use only this model to build our new rankings. 

\subsection{Example 1: Rankings of Scotland in 2013}
\begin{figure}[h]
\begin{center}
\includegraphics[width=\linewidth]{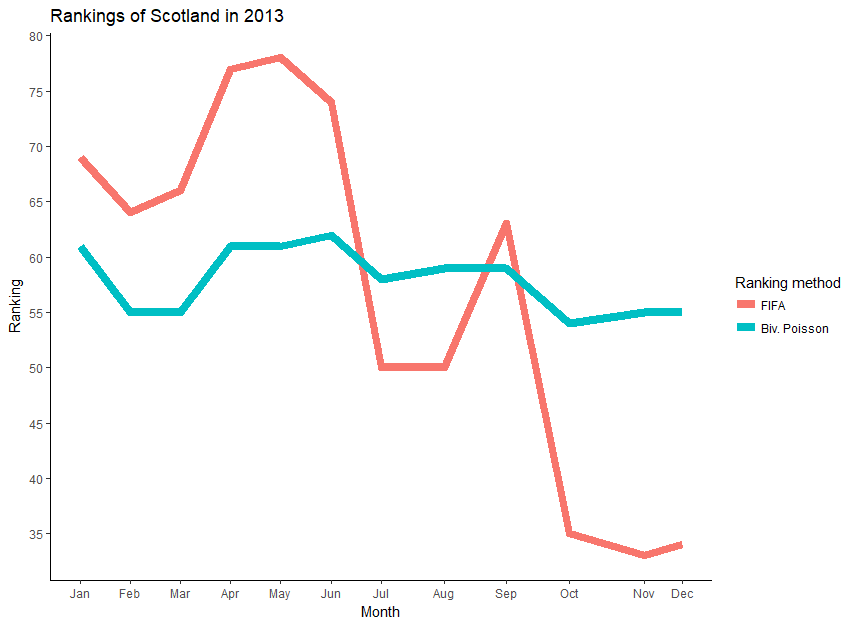}
\end{center}
\caption{Comparison of the evolution of the FIFA ranking of Scotland in 2013 with the evolution based on our proposed ranking method, using the Bivariate Poisson model with 1 strength parameter and Half Period  of 3 years.}
\label{fig:scotland}
\end{figure}

As mentioned in the Introduction, the abrupt decay function of the FIFA ranking has entailed that the ranking of Scotland varied a lot in 2013 over a very short period of time:  ranked $50^{\rm th}$ in August 2013, it dropped to rank 63 in September 2013 before jumping to rank 35 in October 2013. In Figure~\ref{fig:scotland}, we show the variation of Scotland in the FIFA ranking together with its variation in our ranking based on the Bivariate Poisson model with 1 strength parameter and Half Period of 3 years. While both rankings follow the same trend, we clearly see that our ranking method shows less jumps than the FIFA ranking and is much smoother. It thus leads to a more reasonable and stable ranking than the FIFA ranking.

\subsection{Example 2: Drawing for the World Cup 2018}
Another infamous example of the disadvantages of the official FIFA ranking is the position of Poland at the moment of the draw for the 2018 FIFA World Cup (December 1 2017, but the relevant date for the seating was October 16 2017). According to the FIFA ranking of  October 16 2017, Poland was ranked $6^{\rm th}$, and so it was one of the teams in Pot 1, in contrast to e.g. Spain or England which were in Pot 2 due to Russia as host occupying one of the 8 spots in Pot 1. Poland has reached this good position thanks to a very good performance in the World Cup qualifiers and, specifically, by avoiding friendly games during the year before the drawing for the World Cup, since friendly games with their low importance coefficient are very likely to reduce the points underpinning the FIFA ranking. This trick of Poland, who used intelligently the flaws of the FIFA ranking, has  led to unbalanced groups at the World Cup, as for instance strong teams such as Spain and Portugal were together in Group B and  Belgium and England were together in group G.  This raised quite some discussions in the soccer world. In the end Poland was not able to advance to the next stage of the World Cup 2018 competition in its group with Columbia, Japan and Senegal, where Columbia and Japan ended first and second, Poland becoming last. This underlines that the position of Poland was not correct in view of their actual strength.

In Table~\ref{Poland} we compare the official FIFA ranking on October 16 2017 to our ranking based on the Bivariate Poisson model with 1 strength parameter and Half Period of 3 years. In our ranking, Poland occupies only position 15 and would not be in Pot 1. Spain and Colombia would enter  Pot 1 instead of Poland and Portugal. We remark that, in the World Cup 2018, Spain ranked first in their group in front of Portugal while, as mentioned above, Columbia turned out first of Group H while Poland became last. This demonstrates the superiority of our ranking over the FIFA ranking. A further asset is its readability:  one can understand the values of the strength parameters as ratios leading to the average number of goals that one team will score against the other. The same cannot be said about the FIFA points which do not  allow  making predictions.

\begin{table}[h]
\centering 
\caption{\label{Poland} Top of the ranking of the  national teams on 16 October 2017 according to the Bivariate Poisson model with 1 strength parameter and a Half Period of 3 years compared to the Official FIFA/Coca-Cola World Ranking on  16 October 2017.}\vspace{.5cm}
\begin{tabular}{rlr|lc}
\textbf{Position} & \textbf{Team} & \textbf{Strength} & \textbf{Team} & \textbf{Points} \\ 
  \hline
1 & Brazil & 1.753 &    Germany & 1631(1631.05) \\ 
2 & Spain &  1.637 &    Brazil & 1619(1618.63) \\ 
3 & Argentina & 1.628 &    Portugal & 1446(1446.38) \\ 
4 & Germany & 1.624 &    Argentina & 1445(1444.69) \\ 
5 & Colombia &  1.496 &    Belgium & 1333(1332.55) \\ 
6 & Belgium & 1.488 &    Poland & 1323(1322.83) \\ 
7 & France & 1.467 &    France & 1226(1226.29) \\ 
8 & Chile & 1.452 &   Spain & 1218(1217.94)\\ 
9 & Netherlands & 1.424  &   Chile & 1173(1173.14) \\ 
10 & Portugal & 1.417 &   Peru & 1160(1159.94) \\ 
11 & Uruguay & 1.354  &   Switzerland & 1134(1134.5) \\ 
12 & England & 1.341 &   England & 1116(1115.69) \\ 
13 & Peru & 1.303 &   Colombia & 1095(1094.89) \\ 
14 & Poland & 1.277 &   Wales & 1072(1072.45) \\ 
15 & Italy & 1.268  &   Italy & 1066(1065.65) \\ 
16 &  Croatia &1.259 &   Mexico & 1060(1059.6) \\ 
17 & Sweden & 1.253 &   Uruguay & 1034(1033.91) \\ 
18 & Denmark & 1.216 &   Croatia & 1013(1012.81) \\ 
19 & Ecuador & 1.211&   Denmark & 1001(1001.39) \\ 
20 & Switzerland &  1.150 &   Netherlands & 931(931.21)\\ 
\end{tabular}

\end{table}

\subsection{Example 3: Alternative ranking for the Premier League}

\begin{figure}[h]
\begin{center}
\includegraphics[width=\linewidth]{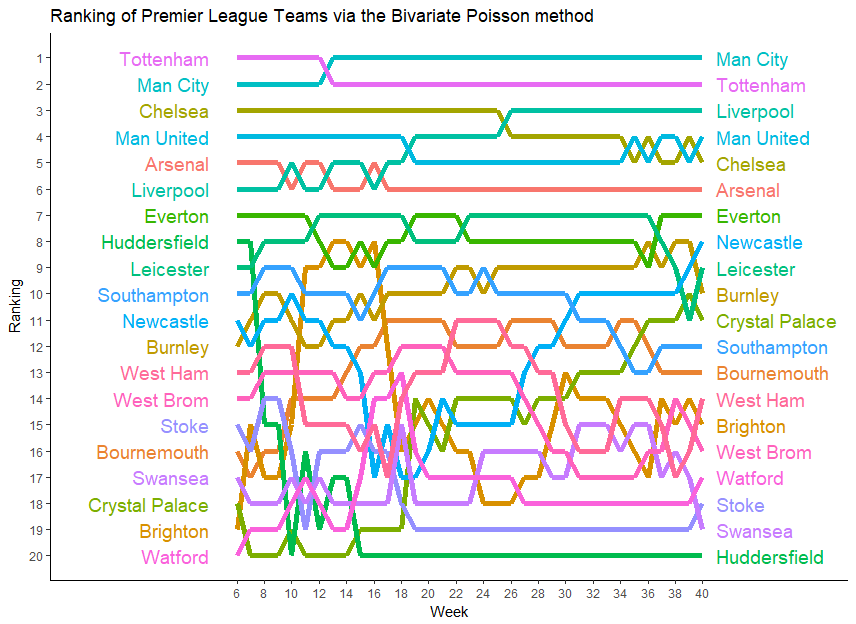}
\includegraphics[width=\linewidth]{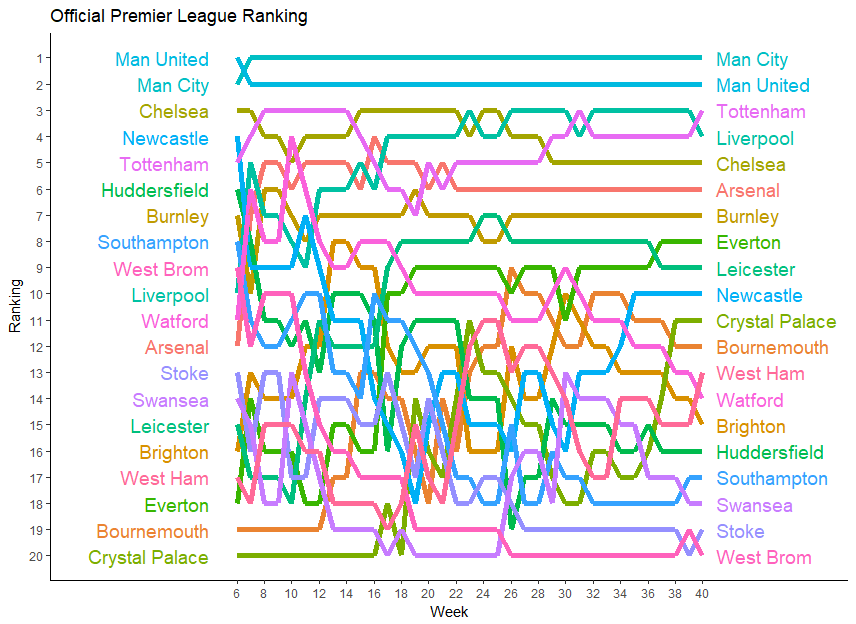}
\end{center}
\caption{Above: Premier League ranking according to  the Bivariate Poisson model with 1 strength parameter and Half Period of 390 days, updated every week, starting from the sixth week since the start of the season. Below: Official Premier League ranking, weekly updated, starting from the sixth week.}
\label{fig:premierleague}
\end{figure}


In Figure \ref{fig:premierleague}, we compare our ranking based on the Bivariate Poisson model with 1 strength parameter and Half Period of 390 days to the official Premier League ranking for the season 2017-2018, leaving out the first five weeks of the season. At first sight, one can see that our proposed ranking is again  smoother than the official ranking, especially in the first part of the season. Besides that, our ranking is constructed in such a way that it does less depend on the game schedules, while the intermediate official rankings are heavily depending on the latter. Indeed, winning against weak teams can rapidly blow up a team's official ranking, while the weakness of the opponents will less increase that team's strength in our ranking which takes the opponent strength into account. Furthermore,  the postponing of matches may even entail that at a certain moment some teams have played more games than others, which of course results in an official ranking that is in favour of the teams which have played more games at that time, a feature that is avoided in our ranking. 

Coming back to the example of Huddersfield Town, mentioned in the Introduction, we can see that our ranking was able to detect Huddersfield as one of the weakest teams in the Premier League after 15 weeks, while their official ranking was still high thanks to their good start of the season. Thus our ranking fulfills its purpose: it reflects well a team's current strength.


\section{Conclusion and outlook}\label{sec:conclu}

We have compared 10 different statistical strength-based models according to their potential to serve as rankings reflecting a team's current strength. Our analysis clearly demonstrates that Poisson models outperform Thurstone-Mosteller and Bradley-Terry type models, and that the best models are those that assign the fewest parameters to teams. Both at domestic team level and national team level, the Bivariate Poisson model with one strength parameter per team was found to be the best in terms of the RPS  criterion. However, the difference between that model and the Independent Poisson with one strength parameter is very small, which is explained by the fact that the covariance in the Bivariate Poisson model is close to zero. This is well in line with recent findings of \cite{Groll2017} who used the same Bivariate Poisson model in a regression context. Applying it to the European Championships 2004-2012, they got a  covariance parameter close to zero.

The time depreciation effect in all models considered in the present paper allows taking into account the moment in time when a match was played and gives more weight to more recent matches. An alternative approach to address the problem of giving more weight to recent matches consists in using dynamic time series models. Such dynamic models, based also on  Poisson distributions, were proposed in \cite{PoissonTimeEffect}, \cite{koopman2015dynamic} and \cite{angelini2017parx}. In future work we shall investigate in detail the dynamic approach and also compare the resulting models to the Bivariate Poisson model with 1 strength parameter based on the time depreciation approach.


\

\noindent \textbf{ACKNOWLEDGMENTS:}

We wish to thank the Associate Editor as well as two anonymous referees for useful comments that led to a clear improvement of our paper.

\

\bibliography{LVdWVE.bib}

\end{document}